\documentstyle[12pt]{article} \textwidth 16.4cm \oddsidemargin 2.5cm
 \advance\oddsidemargin by -1in \evensidemargin 0.0cm \advance\evensidemargin
 by -1in \marginparwidth 1.9cm \marginparsep 0.4cm \marginparpush 0.4cm
 \newcommand{\la}{\langle} \newcommand{\ra}{\rangle} \newcommand\noi{\noindent}
 \newcommand\beq{\begin{equation}} \newcommand\eeq{\end{equation}}
 \newcommand\beqn{\begin{eqnarray}} \newcommand\eeqn{\end{eqnarray}}

\begin{document}

\vspace*{3cm}

\centerline{\Large \bf Charmonium Suppression in Heavy Ion Collisions}

\medskip \centerline{\Large \bf by Prompt Gluons}

\vspace{.5cm}
\begin{center}
 {\large J\"org~H\"ufner$^{1,2}$ and Boris~Z.~Kopeliovich$^{2,3}$}\\ \medskip
{\sl $^1$Institut f\"ur Theoretische Physik der Universit\"at ,\\
Philosophenweg 19, 69120 Heidelberg, Germany}\\

{\sl $^2$ Max-Planck Institut f\"ur Kernphysik, Postfach 103980,
69029 Heidelberg, Germany}\\


{\sl $^3$ Joint Institute for Nuclear Research, Dubna, 141980 Moscow Region,
Russia}

\end{center}

\vspace{.5cm}
\begin{abstract}

In relativistic heavy ion collisions, also the bremsstrahlung of
gluons in the fragmentation regions of the nuclei
suppresses the produced charmonium states.  In the energy range of the
SPS, the radiation of semi-hard gluons
occurs in the Bethe-Heitler regime and the density of gluons
and therefore the suppression goes like $(AB)^{1\over3}$, where
$A$ and $B$ are the nucleon numbers of the projectile and target nuclei.  In
contrast, the suppression via collisions with nucleons is proportional to
$(A^{1\over3} + B^{1\over3})$. Parameter free perturbative QCD calculations are
in a good agreement with the data on $J/\Psi$ and $\Psi'$ suppression in heavy
ion collisions at SPS CERN.  At higher energies (RHIC, LHC) the number of
gluons which are able to break-up the charmonium substantially decreases and
the additional suppression is expected to vanish.

\end{abstract}



\vspace{1cm} \noi {\large\bf Challenges in charmonium production off nuclei}

Charmonium production in heavy ion collisions is considered as one of several
sensitive probes for the formation of a quark-gluon plasma \cite{satz}.  This
idea has led the NA38 and NA50 collaborations at SPS CERN to perform a series
of measurements of nuclear suppression for charmonium production in $p\,A$ and
$A\,B$ collisions.  We recall some remarkable observations (see the collection
of data in \cite{lourenco}):

(i) If $J/\Psi$ suppression in proton-nucleus collisions is treated as a final
state absorption phenomenon one extracts an absorption cross section
$\sigma_{in}(J/\Psi\,N)\approx 6-7\,mb$ from the data.  At the same time, the
vector dominance model applied to $J/\Psi$ photoproduction data on protons
results in a much smaller value $\sigma_{tot}(J/\Psi\,N)\approx 1.1\,mb$, which
increases to a value of $3-4\,mb$ for a coupled-channel analysis of the
photoproduction data \cite{hk}.  However, a discrepancy of about $3\,mb$ still
remains to be explained. A part of this extra suppression 
of $J/\Psi$ may arise due to a
stronger absorption of $\chi_c$ states which contribute to the production of $
J/\Psi$ about $30\%$
via decay\footnote{We are thankful to Mark Strikman for
this comment}.

(ii) The $\Psi'(2S)$ state is expected to have a radius about twice as large as
the $J/\Psi$ and, therefore, to have a few times stronger final state
interaction.  However, in proton nucleus collisions both $J/\Psi$ and $\Psi'$
are suppressed by nearly the same amount.  This is an effect of the finite
formation time of the $J/\Psi$ and $\Psi'$ and can be quantitatively explained
in the coupled channel approach \cite{hk-prl}.

(iii) However, in $S$-$U$ and $Pb$-$Pb$ collisions $\Psi'$ is much stronger
suppressed than $J/\Psi$.  This is partially an effect of the formation time
together with the inverse kinematics for the projectile nucleus \cite{hk-prl}.
Nevertheless nearly a half of the observed suppression still needs to be
explained.

(iv) Final state absorption on nucleons well describes the data on $J/\Psi$
suppression in $p$-$A$, $O$-$Cu$, $O$-$U$ and $S$-$U$ collisions with light
projectiles. However, it fails in the case of $Pb$-$Pb$ collision where data
show a much stronger suppression of $J/\Psi$. This observation presented at
QM'96 \cite{qm96} and published in \cite{na50} has led to intensive theoretical
work (for a recent review we refer to \cite{kh}).  At present there are two
main schools of thought: A number of authors consider the strong suppression in
the $Pb$-$Pb$ data as a proof for the formation of a QGP, in the spirit of the
original idea by Matsui and Satz \cite{satz}.  Other authors explain the strong
suppression of $J/\Psi$ and $\Psi'$ in $Pb$-$Pb$ by hadronic ''comovers'',
{\it i.e.} via destruction of the charmonium by the hadrons which are produced
in the final state of a nucleus-nucleus collision with the same velocity as the
charmonium.  Both approaches, via QGP or comovers, introduce a number of free
adjustable parameters, and it is only then that they can account for the data.
For this reason it is not possible yet to decide about the mechanism.

In this paper we are able to explain the same set of data by invoking -- an
addition to the normal suppression via collisions with nucleons -- the
charmonium suppression due to gluon bremsstrahlung produced in multiple nucleon
interaction.  Since we deal with semi-hard processes, pQCD is expected to work
reasonably well and no adjustable parameters have to be introduced.

\bigskip \noi {\large\bf Charmonium break-up via interaction with nucleons and
gluons}

\begin{figure}[tbh] \includegraphics{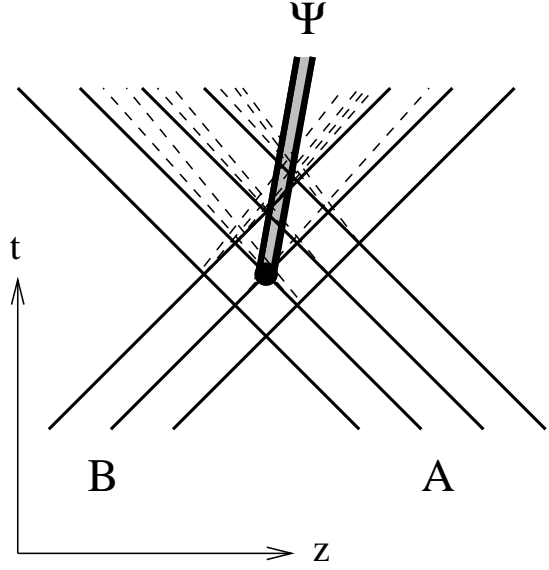}
\begin{center} \vspace{7.5cm} \parbox{13cm} {\caption[Delta]
{\sl A two-dimensional (time - longitudinal coordinate) plot for charmonium
production in a collision of nuclei $A$ and $B$ in the c.m.  of the colliding
nucleons.  The solid and dashed lines show the nucleon and gluon trajectories,
respectively.} \label{fig1}} \end{center} \end{figure}

Final state interaction of a charmonium in nuclear medium which leads to the
nuclear suppression is usually described in terms of collisions of the
charmonium with undisturbed bound nucleons.  This is might not be fully
correct.  A nucleus-nucleus collision is illustrated in Fig.~\ref{fig1} on a
two dimensional (time -- longitudinal coordinate) plot
which corresponds to the $NN$ c.m.  frame.  
One
can see that some of the nucleon trajectories cross the charmonium after they
have interacted with nucleons from another nucleus.  A natural question arises,
whether the charmonium interacts with such debris of a nucleon in the same way
as with an intact one?

To answer this question note, that a small-size $\bar cc$ pair interacts mostly
with only one of the valence quarks in a large-size proton.  This is because of
color screening which cuts off the gluons propagating far away from the
charmonium (the Van der Waals forces are supposed to be cut off by confinement).  
Thus, the charmonium acts like a counter of the number of
constituents in the nucleon. In the constituent quark model we expect
\beq \sigma_{in}(\Psi N)\approx 3\,\sigma_{in}(\Psi q)\ .
\label{2.0}
\eeq
Interacting with the debris of a ''wounded'' nucleon ($N^*$) the charmonium can
find more constituents in it than in the original nucleon, particularly gluons
shown by dashed lines in Fig.~\ref{fig1}, which had a sufficiently short
radiation time.  In this case the $\bar cc$ break-up cross section increases,
\beq \sigma_{in}(\Psi N^*) = \sigma_{in}(\Psi N) + \sigma_{in}(\Psi g)\,\la
n_g\ra\ ,
\label{2.1}
\eeq
where $\la n_g\ra$ is the mean number of gluons radiated in a $NN$ interaction
preceding the collision with the charmonium.

Interaction of the $\bar cc$ pair with ''wounded'' and intact nucleons leads to
the following suppression of charmonium in $AB$ collision,
\beqn &&S^{AB}_{\Psi} = \frac{1}{A\,B}\,\int d^2b\int d^2s
\int\limits_{-\infty}^{\infty} dz_A\,\rho_A(s,z_A)
\int\limits_{-\infty}^{\infty} dz_B\,\rho_B(\vec b-\vec s,z_B)\nonumber\\ &&
\times{\rm exp}\left[-\sigma_{eff}^A(\Psi N)\,
\int\limits_{z_A}^{\infty}dz\,\rho_A(s,z) -\sigma_{eff}^B(\Psi N)\,
\int\limits^{z_B}_{-\infty}dz\,\rho_B(\vec b-\vec s,z) \right]\nonumber\\ &&
\times{\rm exp}\left[-\sigma^A_{eff}(\Psi g)\,\la n_g\ra\, \sigma_{in}(NN)\,
\int\limits_{z_A}^{\infty}dz\,\rho_A(s,z)\,
\int\limits^{\infty}_{-\infty}dz\,\rho_B(\vec b-\vec s,z)\right]
\label{2.2}\\
&& \times{\rm exp}\left[-\sigma^B_{eff}(\Psi g)\,\la n_g\ra\, \sigma_{in}(NN)\,
\int\limits_{-\infty}^{\infty}dz\,\rho_A(s,z)\,
\int\limits^{z_B}_{-\infty}dz\,\rho_B(\vec b-\vec s,z)\right] \nonumber\\ &&
\times{\rm exp}\left[\sigma_{in}(NN)\,\frac{\la n_g\ra}{2}\,
\left[\la\sigma^A_{eff}(\Psi g) + \sigma^B_{eff}(\Psi g) \right]\,
\int\limits_{z_A}^{\infty}dz\,\rho_A(s,z)\,
\int\limits^{z_B}_{-\infty}dz\,\rho_B(\vec b-\vec s,z)\right]\ .
\nonumber \eeqn
Here $\Psi$ denotes either $J/\Psi$ or $\Psi'$.  Effective absorption cross
sections of the $\bar cc$ pair $\sigma_{eff}^{A,B}(\Psi N)$ and
$\sigma_{eff}^{A,B}(\Psi g)$ may be different for nuclei $A$ and $B$ (for
nonzero $x_F$) and depend on where and when the $J/\Psi$ or $\Psi'$ is produced
(see below).

In Eq.~(\ref{2.2}) the coordinates of the production point for the $\bar cc$
pair are assumed to be $(\vec s,z_A)$ and $(\vec b-\vec s,z_B)$ in the nuclei
$A$ and $B$, respectively, and $\vec b$ is the impact parameter for the $AB$
collision; $\rho_{A,B}(\vec r)$ is the density of the nuclei $A$ or $B$.

The first exponential in (\ref{2.2}) accounts for a suppression due to
interaction of the charmonium with the nucleons, both in nuclei $A$ and $B$.
The second and the third exponential factors include suppression of the
charmonium due to interaction with the gluons radiated in the fragmentation
regions of the nuclei $A$ and $B$, respectively.  The last exponential
eliminates a double counting in number of interacting nucleons (assuming for
this purpose that the charmonium is produced with $x_F=0$).

Gluon bremsstrahlung will also affect nuclear suppression in proton-nucleus
collision.  Reducing Fig.~1 to the case of $pA$ collision one can see that some
of gluons radiated by the projectile proton cross the trajectory of the
charmonium.  The expression for nuclear suppression is simpler than
(\ref{2.2}),
\beqn S^{pA}_{\Psi} &=& \frac{1}{A}\,\int d^2b \int\limits_{-\infty}^{\infty}
dz_A\,\rho_A(b,z_A)\, {\rm exp}\left[-\sigma_{eff}^A(\Psi N)\,
\int\limits_{z_A}^{\infty}dz\,\rho_A(b,z) \right]\nonumber\\ && \times{\rm
exp}\left[-\sigma^A_{eff}(\Psi g)\,\la n_g\ra\,
\int\limits_{z_A}^{\infty}dz\,\rho_A(b,z)\right]\ ,
\label{2.3}
\eeqn
Here we neglect the possible effect of dilution of the gluon density due to
transverse motion of the gluons, since for the time interval less than one Fermi
most of them stay within the nucleon size in transverse plane.

In order to evaluate expressions (\ref{2.2}) - (\ref{2.3}), we have to discuss
the input parameters, the mean number of produced gluons $\la n_g\ra$, the
charmonium break-up cross sections $\sigma_{in}(\Psi N)$ and $\sigma_{in}(\Psi
g)$ which are effective cross sections because of the formation time effects.
All the quantities can be estimated fairly reliably.  We concentrate on the
kinematics of SPS which corresponds to about $10\,GeV$ per nucleon in c.m.
frame.  All the longitudinal distances in the nuclei are contracted by a factor
of $\gamma=10$.

\bigskip \noi {\large\bf Mean number \boldmath$\la n_g\ra$ of radiated gluons}

A quark in a projectile nucleon radiates only if interacts with a target
nucleon \cite{gb}.  The mean free path for a quark in nuclear medium is three
times longer than for a nucleon, $\lambda_q\approx 6\,fm$.  One has a maximal
number of produced gluons, namely, the transversal density of gluons
proportional to $(AB)^{1\over 3}$, if the gluons are radiated
independently in each $NN$ interaction in Fig.~\ref{fig1}.  This is possible
only if the radiation happens in the Bethe-Heitler regime, {\it i.e.} if the
formation length $l^g_f$ of the radiation in the c.m.  frame does not exceed
the mean free path of a quark $\Delta z=\lambda_q/\gamma\approx 0.6\,fm$:
\beq l^g_f = \frac{2\,E_q\,\alpha(1-\alpha)} {\alpha^2m_q^2 + k^2} \leq \Delta
z\ ,
\label{3.1}
\eeq
where $E_q$ and $m_q$ are the initial energy and the mass of the radiating
quark; $\alpha$ and $\vec k$ are the fraction of the quark light-cone momentum
and the transverse momentum carried by the gluon.

When the condition (\ref{3.1}) is violated, the Landau-Pomeranchuk effect
suppresses gluon radiation, besides, most of gluons are formed too late to
cross the charmonium trajectory in Fig.~\ref{fig1}.

Thus, the mean number of gluons radiated in a $NN$ collision is given by the
integral over the gluon radiation spectrum weighted by a step-function which
requires the radiation to be in the Bethe-Heitler regime,
\beq \la n_g\ra =\frac{3}{\sigma_{in}(NN)}\, \int\limits_{k^2_{min}}^{\infty}
dk^2 \int\limits_{\alpha_{min}}^1 d\alpha\, \frac{d\sigma(qN\to gX)}
{d\alpha\,dk^2}\, \Theta(\Delta z - l^g_f)\ .
\label{3.2}
\eeq
Here $\alpha_{min}=(\omega_{min}+\sqrt{\omega_{min}^2 - k^2})/(2E_q)$ if
$k < \omega_{min}$, otherwise 
$\alpha_{min}=k/(2E_q)$. The
choice of the soft limit $k_{min}$ has practically no influence on $\la n_g\ra$
because the step function in (\ref{3.2}) cuts off the gluons with small
transverse momenta.  Those which contribute have $k > 0.3-0.5\,GeV$. These are
semi-hard gluons which do not overlap with the soft gluonic field in a
color-flux tube (string).

The cross section of gluon radiation as function of $\alpha$ and $\vec k$
integrated over the final transverse momentum of the quark is calculated in
\cite{kst},
\beq \frac{d\sigma(qN\to gX)} {d\alpha\,dk^2} =
\frac{3\,\alpha_s(k^2)\,C}{\pi}\ \frac{2\,m_q^2\,\alpha^4\,k^2 +
\left[1+(1-\alpha)^2\right](k^4+\alpha^4m_q^4)} {(k^2+\alpha^2m_q^2)^4}\,
\left[\alpha + {9\over 4}\,\frac{1-\alpha}{\alpha}\right]
\label{3.3}
\eeq
Here $\alpha_s(k^2)$ is the QCD running coupling; $C$ is the factor for the
dipole approximation for the cross section of a $\bar qq$ pair with a nucleon,
$\sigma^{\bar qq}(r_T)\approx C\,r_T^2$, where $r_T$ is the $\bar qq$
transverse separation \cite{zkl,kst}.  PQCD predicts $C\approx 3$.  Note that
(\ref{3.3}) is supposed to work well even for radiation of gluons with rather
small $k$ \cite{kst}.

We are left with only one parameter $\omega_{min}$ in (\ref{3.2}) which brings
the main uncertainty in the value of $\la n_g\ra$.  In accordance with the
consideration in next section we try two values $\omega_{min}= 0.5$ and
$1\,GeV$, which result in $\la n_g\ra=0.69$ and $0.25$, respectively.

Note that the radiation spectrum steeply grows with energy \cite{kst} since we
select gluons with relatively large $k_T$. According to the analysis
\cite{kp,kst} of HERA data for the proton structure function we expect energy
dependence $d n_g/d\alpha dk^2 \propto (s/s_0)^{0.2}$.  However, the
restriction for the formation length imposed by the step function in
(\ref{3.2}) suppresses $\la n_g\ra$ much more at high energy because of Lorentz
contraction of the nuclei.

We compare $\la n_g\ra$ for $AB$ collision calculated for $\omega_{min}=0.5\,GeV$ 
at SPS, RHIC and LHC
energies including also the energy growth of the gluon spectrum.
\beq \la n_g\ra = \ \left\{\begin{array}{cc}6.9\times 10^{-1}\ \ \ &(SPS,\
\sqrt{s}=20\,GeV)\\ 6.9\times 10^{-3}\ \ \ &(RHIC,\ \sqrt{s}=200\,GeV)\\ 1.2\times
10^{-3}\ \ \ &(LHC,\ \sqrt{s}=1200\,GeV)\end{array}\right.\ .
\label{3.5}
\eeq
In the rest frame of either of the colliding nuclei the main fraction of gluons
are radiated at high energies later than the charmonium is produced.

\bigskip \noi {\large\bf The inelastic gluon - charmonium cross section
\boldmath$\sigma_{in}(\Psi g)$}

The choice of $\omega_{min}$ deserves a special discussion since it is related
to the problem of the energy dependence of $J/\Psi-g$ break-up cross section
$\sigma_{in}(J/\Psi)$.  At high energies it is dominated by gluonic exchange
in $t$-channel and slightly grows with energy \cite{hk}.  According to
the previous discussion and Eq.~(\ref{2.0}) the inelastic cross section
with a quark is known. The one with a gluon differs by a color factor
$9/4$. Therefore, we expect
\beq \sigma_{in}(J/\Psi g) \approx {3\over4}\, \sigma_{in}(J/\Psi N)\ .
\label{4.1}
\eeq

In the energy range of interest a coupled-channel analysis of $J/\Psi$
photoproduction data gives $\sigma_{in}(J/\Psi N)\approx
4\,mb$. Correspondingly, $\sigma_{in}(J/\Psi g)\approx 3\,mb$. For $\psi'(2S)$
we expect a larger cross section. If the cross section is proportional to $\la
r^2\ra$ of the charmonium state, one has a factor of $7/3$ in an oscillator
model, or a factor of $4$ for $\la r^2\ra$ calculated from realistic wave
functions \cite{buch}.

One could expect a vanishing $\sigma_{in}(\Psi N)$ down to the
threshold energy which is about $0.6\,GeV$ lower for $\Psi'$ than for $J/\Psi$.
However, since radiation time is of the order of nuclear radius, the gluons
cross the charmonium trajectory in Fig.~1 at about the same time as the wounded
nucleons, {\it i.e.}, when the charmonium is not yet formed.  One cannot
require energy conservation to hold with accuracy better than $\Delta E
\approx 1/\Delta t$ where $\Delta t$ is the time interval between the
production moment of $\bar cc$ (the production/coherence time is quite short at
the SPS energy) and its crossing with a gluon.  The mean time of crossing in
Fig.~1 is $\Delta t\approx (1/3)R_A/\gamma\approx 1\,GeV^{-1}$.  Therefore, one
can disregard energy conservation, binding energy and the difference in
threshold energy between $J/\Psi$ and $\Psi'$ with accuracy 
$\Delta E\sim 1\,GeV$.  For the same reason, the energy
threshold has no influence on $\sigma_{in}(\Psi\,g)$.

There is also another mechanism of charmonium break up by gluons via direct
absorption of the gluon \cite{shuryak} which has a pick about
$\sigma_{in}(\Psi\,g)\approx 5\,mb$ near the threshold with a width of the order
of binding energy and falls steeply at higher energies.
In this case one should replace the binding energy by the $\Delta E$.

To simplify calculations we fix $\sigma_{in}(J/\Psi g) \approx 3\,mb$
at $\omega > \omega_{min}$.  We try $\omega_{min} =0.5$ and $1\,GeV$ in
accordance with the scale imposed by the estimated uncertainty in energy.

\bigskip \noi {\large\bf  Effective absorption cross sections}

The produced $\bar cc$ colorless pair propagates through the nucleus with a
smaller size and a smaller absorption than that for either of $J/\Psi$ and
$\Psi'$, unless time exceeds the formation length $l^{\Psi}_f$ for the
charmonium wave function \cite{kz,hk-prl},
\beq l^{\Psi}_f=\frac{2E_{\Psi}}{M^2_{\Psi'} - M^2_{J/\Psi}}\ .
\label{5.1}
\eeq
Here $E_{\Psi}$ is the charmonium energy in the rest frame of the nucleus.  The
transition between short and long times is controlled by the nuclear formfactor
\cite{hk-prl}, which  for the charmonium production
point with impact parameter $\vec s$  reads,
\beq F_A(q_L,s)= \frac{1}{T(s)}\, \int\limits_{-\infty}^{\infty}dz\,e^{iq_Lz}\,
\rho_A(s,z)\ ,
\label{5.2}
\eeq
where the longitudinal momentum transfer is related to (\ref{5.1}),
$q_L=1/l^{\Psi}_f$, and $T(b)=\int_{-\infty}^{\infty}dz\,\rho_A(b,z)$ is the
nuclear thickness function.

The break-up of the colorless $\bar cc$ pair during its evolution is described
by the effective cross section \cite{hk-prl},
\beqn \sigma_{eff}^A(J/\Psi N)&=& \sigma_{in}(J/\Psi N)\,
\Bigl[1+\epsilon\,R\,F^2_A(q_L^A,s)\Bigr]
\label{5.3}\\
\sigma_{eff}^A(\Psi' N)&=& \sigma_{in}(\Psi' N)\,
\Bigl[1+\frac{\epsilon}{r\,R}\,F^2_A(q_L^A,s)\Bigr]\ .
\label{5.4}
\eeqn
The parameters here are defined in \cite{hk-prl}.  We use the values given by
the harmonic oscillator model, $r=\la\Psi'|\hat\sigma|\Psi'\ra/ \la
J/\Psi|\hat\sigma|J/\Psi\ra = 7/3$, $\epsilon = \la\Psi'|\hat\sigma|J/\Psi\ra/
\la J/\Psi|\hat\sigma|J/\Psi\ra = -\sqrt{2/3}$.  The relative production rate
of $\Psi'$ to $J/\Psi$ in a $NN$ collision is known from experiment, $R\approx
0.4$.

We expect different effective cross sections for absorption in the nuclei $A$
and $B$, since the charmonium with $\la x_F\ra = 0.15$ (NA38, NA50) propagates
through these nuclei with different energies in their rest frames, {\it i.e.}
$q_L^A$ and $q_L^B$ are different.  In an $AB$ collision with
$E_{lab}=158\,GeV/A$and $x_F=0.15$ one has $l_f^A=2\,fm$ and $l_F^B=4\,fm$ in
the rest system of the projectile (A), target (B), respectively.  The values
for the effective cross sections in a $Pb$-$Pb$ collision are
(Eqs.~(\ref{5.3})-(\ref{5.4})) $\sigma_{eff}^A(J/\Psi N) =3.9\,mb$,
$\sigma_{eff}^B(J/\Psi N)=3.3\,mb$, $\sigma_{eff}^A(\Psi' N)=8.6\,mb$ and
$\sigma_{eff}^B(\Psi' N)=4.8\,mb$ (for $\sigma_{in}(J/\Psi N)=4\,mb$ and
$\sigma_{in}(\Psi' N)= 9.3\,mb$)

Due to the uncertainty for the radiation time one can assume that the gluons
radiated by the ''wounded'' nucleons cross the $\Psi$ trajectory at about the
same points as the nucleons.  Therefore, we can use the same form of the
effective cross sections (\ref{5.3}) - (\ref{5.4}) for interaction with gluons.

\bigskip \noi {\large\bf Results of calculations}

To simplify the calculations we use a constant nuclear density
$\rho(r)=\rho_0\,\Theta(R_A-r)$ with $\rho_0= 0.16\,fm^{-3}$, except the
nuclear formfactor where a realistic density is used.  We calculate (\ref{2.2})
and (\ref{2.3}) using the values of the parameters fixed in previous sections.
We use two values of $\la n_g\ra = 0.25$ and $0.5$ to characterize of the
uncertainty in our calculations.

The results are plotted in Fig.~2a for $J/\Psi$ and Fig.~2b for $\Psi'$ as
function of $A\times B$ together with available experimental data for $pA$ and
$AB$ collisions.
\begin{figure}[tbh]
\includegraphics{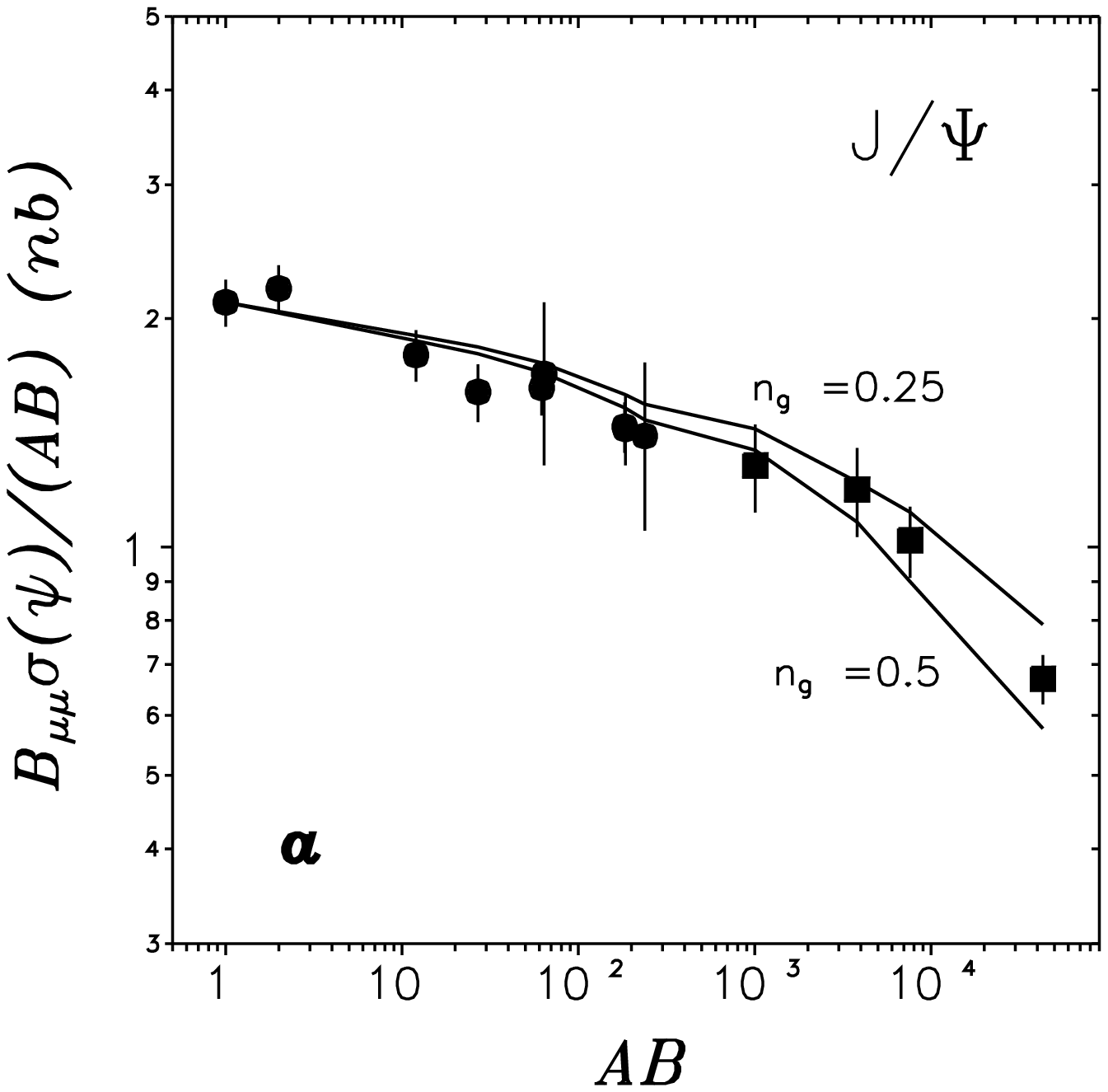}  \includegraphics{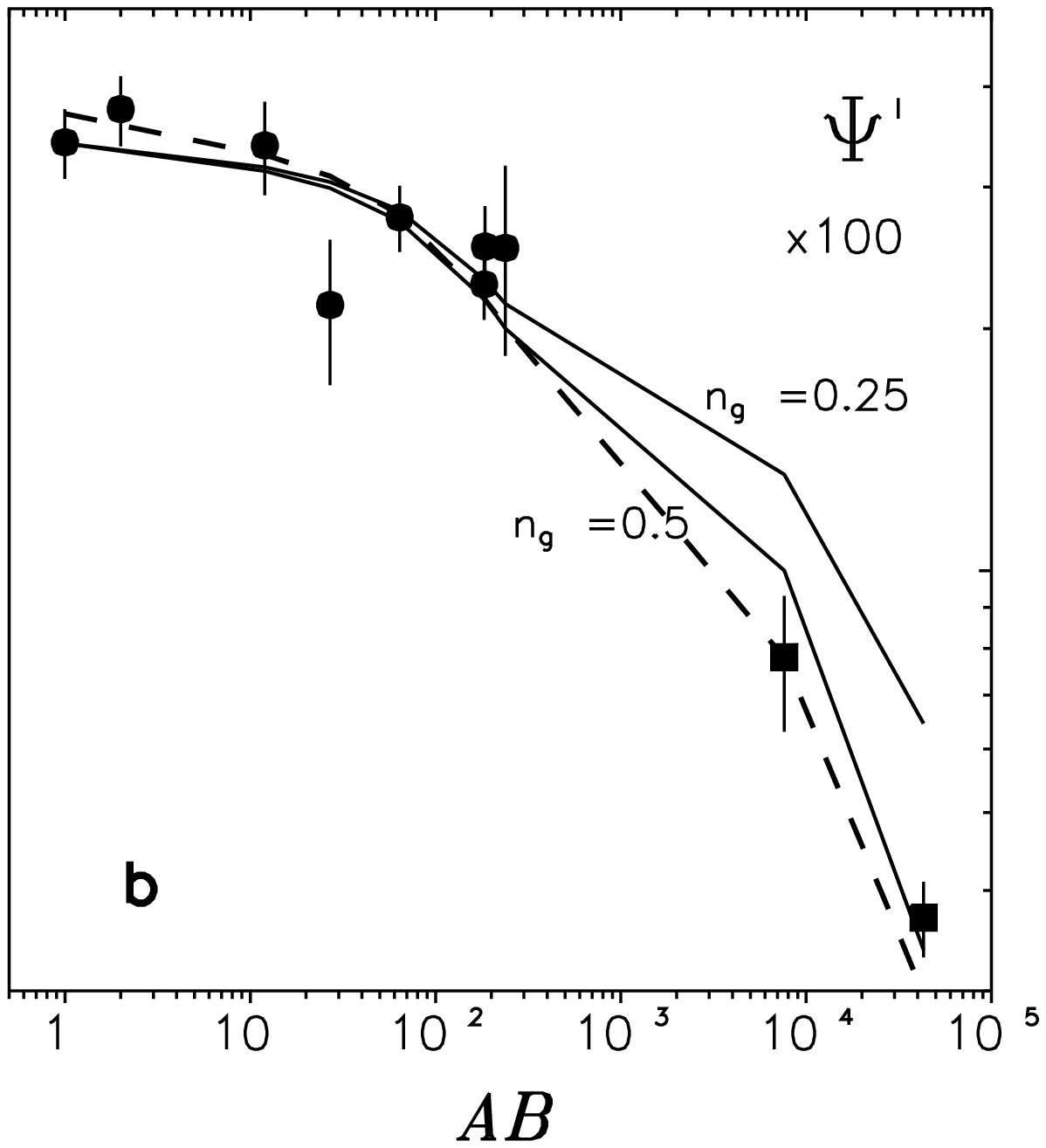}
\begin{center}
\vspace{8.5cm} \parbox{13cm} {\caption[Delta] {\sl {\bf (a)} Nuclear
suppression of $J/\Psi$ production in $pA$ and $AB$ collisions as function of
the product $A\times B$.  The two curves correspond to $\la n_g\ra=0.25$ and
$0.5$ The lines connect the point calculated for $A\times B$ corresponding to
each experimental point. {\bf (b)}
The same as on (a), but for $\Psi'$ production rescaled by factor $10^{-2}$.
The circles and squares correspond to $pA$ and $AB$ data, respectively.  The
data points are from \cite{lourenco,qm96,na50}.}
\label{fig2}}
\end{center}
\end{figure}
We reproduce the data for $J/\Psi$ suppression quite well, including the
$S$-$U$ and $Pb$-$Pb$ points.  We obtained a reasonable agreement with the data
for $\Psi'$ suppression as well, although the curves seem to be a bit too high.
In this respect we should note that the two coupled channel model we use with
oscillator potential might be a poor approximation in the case of $\Psi'$.
Indeed, as was mentioned, the ratio of mean square radiuses of $\Psi'$ to
$J/\Psi$ which is $7/3$ in the oscillator model, is predicted to be nearly $4$
in a more realistic approach.  As an example we tried another value $r=3$ in
(\ref{5.3}) - (\ref{5.4}).  The result is shown by the dashed curve in Fig.~2b
which is in a better agreement with the data.  Our results for $J/\Psi$
suppression are very stable against this modification.

Note that our curves are not straight lines even for $pA$ collisions.  This is
mostly due to the formation time effects, {\it i.e.} is a result of
$A$-dependence of the effective cross sections (\ref{5.3}) - (\ref{5.4}).
Furthermore, we are able to account for the $J/\Psi$ suppression in $pA$
collisions, although we use $\sigma_{in}(J/\Psi N)=4\,mb$. The additional
suppression (even in $pA$) comes from the gluons with absorption cross section
of $3\,mb$.  At larger values $A\times B$ for nuclear collisions 
our curves deviate even more from
straight lines (in agreement with the data!) due to the nonlinear dependence on
nuclear radius, as is brought into (\ref{3.2}) by the gluon radiation, whose
density is proportional to $(AB)^{1/3}$.

\begin{figure}[tbh] 
\includegraphics{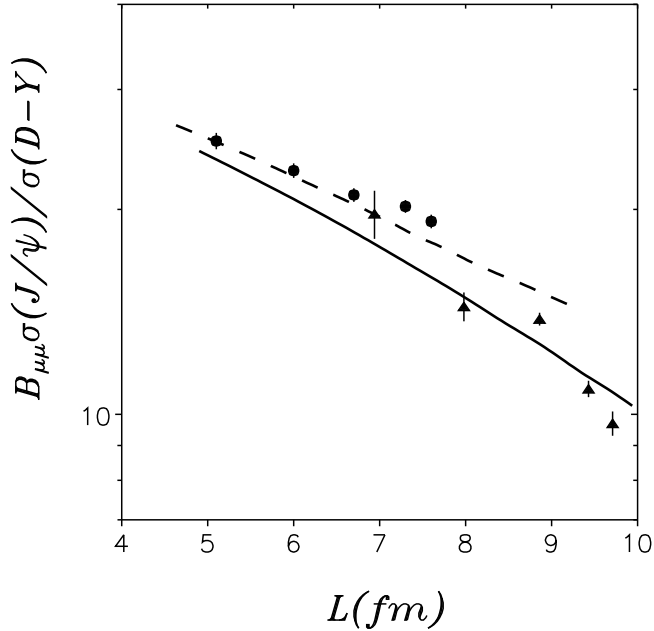}
\begin{center} 
\vspace{8.5cm} 
\parbox{13cm} 
{\caption[Delta]
{\sl Nuclear suppression of $J/\Psi$ relative to 
Drell-Yan lepton pairs as function of the mean length of 
total path of the charmonium in the colliding nuclei.
The dashed (round points) and solid (triangles) 
curves correspond to S-U and Pb-Pb 
collisions, respectively. The data points are from \cite{na50}.} 
\label{fig3}} 
\end{center} 
\end{figure}

Formulas (\ref{2.2}) and (\ref{2.3}) can be also used for calculation of
nuclear suppression of charmonia as a function of impact parameter or of the
mean length $L$ of the total path of $\Psi$ in nuclear matter.  
Our results for S-U and Pb-Pb collisions are plottes in Fig.~\ref{fig3}
by dotted and solid curves respectively. We use $\la n_g\ra=0.5$ at
$200\,GeV$ and $\la n_g\ra=0.57$ at $158\,GeV$ in accordance with 
Eq.~(\ref{3.2}). The overall normalization is free.

Following recommendation of the referee for our paper we compare the
calculations with the published data \cite{na50} depicted in
Fig.~\ref{fig3} by round and triangle points for S-U and Pb-Pb
collisions respectively. However, the
assignment for $L$ corresponding to the measured transverse energy is not
trustable. It is based on an oversimplified model, particularly neglecting
the Landau-Pomeranchuk suppression (at low $k$) and enhancement (at high $k$)
of the gluon radiation spectrum \cite{kst}.  The cascading of gluons radiated in
the Bethe-Heitler regime which increases the transverse energy, is also
neglected. Besides, the normalization on the cross section of the Drell-Yan
reaction brings an additional uncertainty (especially at high energies). 
We plan to do our own calculations for the transverse energy produced in heavy
ion collisions which would allow an unbiased comparison with data.

\bigskip \noi {\large\bf Conclusions}

Charmonium suppression in $pA$ and $AB$ collisions at SPS energies can be
calculated as arising from two sources:

(i) Collisions with nucleons, the cross section for which is the one deduced
from photoproduction data modified because of the formation time effects. The
suppression from this source essentially scales like $A^{1/3} + B^{1/3}$.

(ii) Collisions of charmonia with gluons which have been produced in multiple
$NN$ collisions.  In the Bethe-Heitler regime - good at SPS energies - about
$\la n_g\ra = 0.25\,-\,0.5$ gluons contribute per $NN$ collision.  This
additional effect scales with $(AB)^{1/3}$ and accounts for the ''anomalous''
$J/\Psi$ and $\Psi'$ suppression in nucleus-nucleus collisions.  We conclude
that the results of the NA38/50 experiments do not signal about QGP formation.

Nonetheless, the suggested mechanism leaves room for other contributions to the
charmonium suppression within the uncertainty of calculations. Particularly, 
as was mentioned above, a ''wounded'' nucleon is not a colorless system of partons
any more. Therefore, color screening does not cut off 
(only the charmonium formfactor does) soft gluon exchanges in contrast to 
$\Psi N$ interaction.
However this correction to (\ref{2.0}), which we do not expect to be large, scales with
$(A^{1/3} + B^{1/3})$ and leads to a renormalization of $\sigma_{in}(\Psi N)$.

According to (\ref{3.5}) the suppression of charmonium by gluon radiation
vanishes at high energies due to the formation time effect. We expect
practically no additional suppression by prompt gluons at RHIC, while the
suppression via interaction with ''comovers'' (either QGP or hadrons) increases
with energy.  Therefore, one has less freedom in interpretation of data on
nuclear suppression of charmonium at RHIC.

The main parameter which controls the charmonium suppression via interaction
with prompt gluons in (\ref{2.2}) is the product $\la n_g\ra\sigma_{eff}(\Psi N)$.
  We estimated this factor and allowed it to vary within a factor of 2. We
think that this uncertainty covers other possible corrections, for instance,
the contribution of $\chi_c$ decays to the production of $J/\Psi$.  \medskip

\noi {\bf Acknowledgements}: we are grateful for useful discussions to Sverker
Fredriksson, Yuri Kovchegov and Hans Pirner.  This work was completed when
B.Z.K.  was visiting at the University and INFN Trieste.  He thanks Daniele
Treleani for hospitality and very helpful discussions. The work is partially
supported by a grant 06 HD742 from the BMBW.

\setlength{\baselineskip} {10pt}


\begin{thebibliography}{MMM}

\bibitem{satz} T.~Matsui and H.~Satz, Phys.  Lett.  {\bf B178} (1986) 416

\bibitem{lourenco} C.~Lourenco, in Proc.  of the Quark Matter Conf.  1996,
Nucl.  Phys.  {\bf A610} (1996) 552c

\bibitem{hk} J.~H\"ufner and B.Z.~Kopeliovich, Phys.  Lett.  {\bf B426} (1998)
154

\bibitem{hk-prl} J.~H\"ufner and B.Z.~Kopeliovich, Phys.  Rev.  Lett.  {\bf 76}
(1996) 192

\bibitem{qm96} M.~Gonin, in proc.  of the Quark.  Matter.  Conf.  1996, Nucl.
Phys.  {\bf A610} (1996) 404c

\bibitem{na50} M.C.~Abreu et al., Phys.  Lett. {\bf B410} (1997) 337

\bibitem{kh} D.~Kharzeev, invited talk at the Quark Matter Conf.  1997,
nucl-th/9802037

\bibitem{gb} J.F.~Gunion and G.~Bertsch, Phys. Rev.  {\bf D25} (1982) 746

\bibitem{kst} B.Z.~Kopeliovich, A.~Sch\"afer and A.V.~Tarasov, hep-ph/9808378

\bibitem{zkl} A.B.~Zamolodchikov, B.Z.~Kopeliovich and L.I.~Lapidus, JETP
Lett. {\bf 33} (1981) 595.

\bibitem{kp} B.Z.~Kopeliovich and B.~Povh, hep-ph/9806284

\bibitem{buch} W.~Buchm\"uller and S.-H.~H.~Tye, Phys.  Rev.  {\bf D24} (1981)
132

\bibitem{shuryak} E.V.~Shuryak, Sov. J. Nucl.  Phys. {\bf 28} (1978) 408

\bibitem{kz} B.Z.~Kopeliovich and B.G.~Zakharov: Phys.Rev.  {\bf D44} (1991)
3466.


\end{thebibliography}
\end{document}